\documentclass[a4,10pt]{article}
\PassOptionsToPackage{hyphens}{url}
\usepackage{hyperref}

\usepackage{tikz}
\usepackage{amsmath}

\usepackage{filecontents}

\def\isanonymous{0}

\usepackage[l2tabu,orthodox]{nag}

\usepackage{ifxetex}

\usepackage{microtype}
\usepackage{booktabs,caption}
\usepackage{hyphenat}
\usepackage[flushleft]{threeparttable}

\usepackage{ifthen}
\newcommand{\anonymous}[2]{%
	\ifthenelse{\equal{\isanonymous}{1}}%
	{{#1}}%
	{{#2}}%
}

\usepackage{xcolor}
\definecolor{oxygenorange}{HTML}{FFDD00}
\usepackage[color=oxygenorange]{todonotes}

\ifthenelse{\equal{\isanonymous}{1}}
{
	\newcommand{\rikke}[2][]{}
	\newcommand{\mbabazi}[2][]{}
}{
	\newcommand{\rikke}[2][inline]{\todo[#1]{\textbf{rikke:} #2}\xspace}
	\newcommand{\mbabazi}[2][inline]{\todo[#1]{\textbf{mbabazi:} #2}\xspace}
}

\usepackage{amsmath,amsfonts}  
\usepackage{xspace}
\usepackage{cleveref}
\usepackage{rotating}

\usepackage[lambda,landau,operators,probability,sets,logic,complexity,asymptotics]{cryptocode}

\usepackage{booktabs}  
\usepackage{comment}
\usepackage{enumitem}

\usepackage{navigator}
\usepackage{url}
\usepackage{graphicx}
\usepackage[export]{adjustbox}
\usepackage{float}
\usepackage{multirow}
\usepackage{longtable}
\usepackage{lscape}
\usepackage{array}
\newcolumntype{L}{>{\raggedright\arraybackslash}p{3cm}}
\usepackage{pdfpages}
\usepackage{subcaption}   
\usepackage{tikz,pgfplots,pgfplotstable}
\usepgfplotslibrary{statistics}
\usetikzlibrary{calc}
\usetikzlibrary{arrows}
\usetikzlibrary{positioning}

\pgfplotsset{
	tick label style={font=\small},
	label style={font=\small},
	legend style={font=\small, cells={anchor=west}}
}

\definecolor{DarkPurple}{HTML}{332288}
\definecolor{DarkBlue}{HTML}{6699CC}
\definecolor{LightBlue}{HTML}{88CCEE}
\definecolor{DarkGreen}{HTML}{117733}
\definecolor{DarkRed}{HTML}{661100}
\definecolor{LightRed}{HTML}{CC6677}
\definecolor{LightPink}{HTML}{AA4466}
\definecolor{DarkPink}{HTML}{882255}
\definecolor{LightPurple}{HTML}{AA4499}

\definecolor{DarkBrown}{HTML}{604c38}
\definecolor{DarkTeal}{HTML}{23373b}
\definecolor{LightBrown}{HTML}{EB811B}
\definecolor{LightGreen}{HTML}{14B03D}

\usepackage{listings}
\lstdefinelanguage{Sage}[]{Python}{morekeywords={True,False,sage,cdef,cpdef,ctypedef,self},sensitive=true}
\begin{document}
	
	\title{Information Security in Small-Scale Protests: \\ Surveillance of Ugandan Anti-EACOP Protesters}
	\anonymous{
		\author{}
	}{
		\author{
			{\rm Ntezi Mbabazi}\\
			Royal Holloway, University of London\\
			ntezi.mbabazi.2023@live.rhul.ac.uk
			\and
			{\rm Rikke Bjerg Jensen}\\
			Royal Holloway, University of London\\
			rikke.jensen@rhul.ac.uk			
		}
		
	}
	
	\date{}
	
	\clubpenalty=1
	\displaywidowpenalty=1
	\widowpenalty=1
	
	\maketitle

\begin{abstract}
We examine the information security practices of Ugandan climate activists protesting the development of the East African Crude Oil Pipeline (EACOP). We conducted five-week fieldwork in Kampala, Uganda, which included interviews with 13 anti-EACOP activists. Through an inductive analysis, we report on the complexities faced by small groups of predominantly student protesters as they covertly organise small-scale anti-EACOP protests within a context marked by state surveillance and repression. Our study points to a multi-layered adversarial landscape, where participants' experiences of direct threats, including arrests and information compromise, and their fears of abduction, shaped their security practices. These practices were rooted in autonomous decision-making within groups. We present a grounded understanding of how participants' need to protect information for their own security, as well as that of others, permeated their lives, leading them to adjust day-to-day aspects of their device management, communication, accommodation, transport and social relations as deliberate tactics to mitigate surveillance.  
\end{abstract} 

\section{Introduction}\label{sec:introduction}

\noindent Activism has long been a tool for different groups to challenge entrenched power structures~\cite{FuentesGeorge2021}. For climate activists, protests have become a critical means of drawing attention to the effects of climate change and policies that activists argue exacerbate such effects~\cite{DAMOAH2023}. In Uganda, climate protests exist within a restrictive political landscape under Museveni's 40-year presidency that is increasingly violent in suppressing dissent~\cite{abrahamsen2021uganda,wilkins2021briefing}. While Uganda's 1995 constitution guarantees freedom of assembly and expression, the reality for activists and protesters is one of state surveillance and violent confrontations with Ugandan security forces, while restrictive laws and a history of crackdowns shape the protest environment~\cite{chibita2016digital,selnes2020activism}. Protests are characterised by excessive use of force from security forces who often disperse protesters using tear gas, water cannons, arrests and, at times, live ammunition~\cite{kagoro2023status,larok2017uganda}. Despite this environment, protests are used to express discontent with social, economic and political issues, such as corruption~\cite{marchtoparliamentprotest,ugandacracksdown}, elections~\cite{kagoro2024uganda}, service delivery~\cite{mbazira2013service}, unemployment~\cite{van2019protest} and climate change~\cite{arrested2024}.

Our study examines the security practices of Ugandan climate activists, many of whom are financially-constrained university students covertly organising small-scale protests against the development of the East African Crude Oil Pipeline (EACOP)~\cite{businesshumanrightsUgandaUniversity,arrested2024}. Operating under intense state surveillance and political pressure, these activists navigate daily threats that shape both their safety and resistance, prompting them to develop discreet and adaptive practices to protect themselves and sustain their protests. Such practices, which are deeply intertwined with their social lives, are driven by a necessity to protect information about themselves and their activism from immediate adversaries such as Ugandan security forces and foreign commercial interests.   

EACOP is designed to transport crude oil from Uganda's oil fields to Tanzania's Chongoleani Peninsula near Tanga port for export. Those opposing EACOP highlight environmental and climate risks, with resistance in both countries~\cite{heede2022east}. Our focus lies in the Ugandan context due to the significantly higher frequency of protests compared to Tanzania. We explore the distinct, often subtle, information security strategies that protesters employ to secure themselves and their justification for these strategies. Our work considers the threats experienced by Ugandan anti-EACOP protesters, their fears and how they organise themselves in small, interlocking groups. We thus contribute to an emerging, yet nascent, body of qualitative information security scholarship that has explored the security needs of activists, e.g.~\cite{USENIX:ABJM21} on the Anti-ELAB protests in Hong Kong,~\cite{SP:DSKB21} on the Sudanese revolution,~\cite{CHI:Guntrum24} on the anti-military protests in Myanmar and~\cite{CHI:LHKZH20} on activism among transgender people.

\paragraph{Contributions.} The central objective of our work was to establish and understand the information security concerns and goals of anti-EACOP protesters in Uganda. We achieved this through five weeks of fieldwork in Kampala which enabled us to conduct \emph{in-situ} interviews with 13 activists, 11 of whom were actively involved in planning and staging anti-EACOP protests and two activists providing logistical and legal support. Being \emph{in situ} enabled contextual understanding of their information security concerns. Our work speaks to a highly adversarial landscape for anti-EACOP protesters, dominated by the Ugandan security forces and foreign commercial interests. Grounding empirical findings, our work makes the following contributions to information security scholarship:

\begin{itemize}
	\setlength\itemsep{0em}
	\item \emph{The small scale of anti-EACOP made protesters highly visible to Ugandan security forces}. This reality necessitated particular information protection strategies not observed in existing security literature focusing on larger-scale protests.
	\item \emph{Surveillance tactics by Ugandan security forces permeated anti-EACOP protesters' lives, necessitating information protection within and beyond activist circles}. This led participants to develop extensive, often individualised, security practices: living in temporary accommodation to avoid having a known home address, changing daily routines and transport modes, and providing false personal details at public facilities and during arrests.
	\item \emph{Compromise took the form of (frequent) arrests}. Arrests were expected and considered a critical point of compromise, while fears of abduction and torture drove participants' security practices. Considerations of compromise meant participants compartmentalised information access within protest groups to protect information about past and future activities, and group membership that could lead to further compromises.
	\item \emph{Participants did not rely on the security of a piece of technology for their protection, citing data cost as a limiting information security factor}, but developed socially-grounded, interpersonal and context-specific practices to protect information about themselves and their activism.
	\item Protest groups comprised loosely connected structures central to the organising of protest activities, yet \emph{individual autonomy prevailed over collective decision-making}. 
\end{itemize}

\noindent We develop these contributions in~\Cref{sec:discussion}. 

\section{Preliminaries on Anti-EACOP Protests}\label{sec:activism-uganda} 

\noindent EACOP is developed by TotalEnergies EP Uganda, a subsidiary of French TotalEnergies, the Ugandan National Oil Company (UNOC), Tanzania Petroleum Development Corporation (TPDC), and China National Offshore Oil Corporation (CNOOC). The 1,443km EACOP ~\cite{mukanza2023conflict} travels along the Albertine Rift, an ecologically rich area~\cite{plumptre2007biodiversity}. Critics argue that the project challenges Uganda's Paris Agreement commitments to reduce emissions by 22\% by 2030~\cite{fidhoil2020}, and that its environmental and social impact assessments underestimate the project's cumulative impacts on communities and ecosystems~\cite{fidhoil2020,heede2022east}. Studies point to ecological risks such as habitat destruction, biodiversity loss, water contamination and deforestation, particularly affecting marginalised groups~\cite{huxham2020understanding,nathan2021assessing,ogwang2018impacts}. Lawsuits against EACOP in East Africa and France have cited environmental risks as their main grounds, but have typically been dismissed over jurisdictional and procedural issues~\cite{africanParisCourt}. 

Ugandan climate activists employ various tactics, including protests, legal actions and environmental clean-ups as forms of resistance. In September 2022, the European Union (EU) Parliament passed a resolution to stop the EACOP project, citing human-rights and environmental concerns. This resolution elicited mixed reactions in Uganda and triggered the country's first major anti-EACOP protest in which nine students were arrested while marching in support of the resolution~\cite{businesshumanrightsUgandaUniversity}. The Ugandan Government dismissed the EU's stance as motivated by racism, sabotage and foreign interference~\cite{EUParliament}, echoing comments made by President Museveni about ``foreign interests'' sabotaging the Ugandan oil sector to prevent the country from attaining financial freedom~\cite{voanewsUgandaPresident}. The deputy speaker of the Ugandan Parliament described the resolution as ``the highest form of economic sabotage against developing countries''~\cite{ParlaimentslamsEU2022}. Thus, Ugandan climate activists also risk being labelled `economic saboteurs' if their protests oppose State economic interests~\cite{lyons2017reforming}, a designation that is used to justify intense scrutiny of their activities, targeting and interrogation by the State.

Ugandan anti-EACOP protesters have staged several protests despite facing excessive force and arrests by the Ugandan security forces. For example, 20 protesters were arrested and detained in November 2023, seven in May 2024 and 30 in June 2024~\cite{frontlinedefendersSevenEnvironmental,ugandacracksdown,monitor2024arrest}. The frequency of protests and scale of arrests grew in 2023 and 2024, with two protests being held in August 2024 (during the fieldwork for our study), in which 47 and 18 protesters were arrested, respectively~\cite{18arrested2024,arrested2024}.

\section{Related Work}\label{sec:related-work}

\noindent In~\Cref{sec:rw-state-surveillance} we consider practices of state surveillance in Uganda, and in activism and surveillance more broadly. Our research is situated in prior qualitative scholarship on the information security of protesters (\Cref{sec:rw-protesters-security-practices}), while we identity specific research gaps that our work aims to address.

\subsection{State Surveillance and Activism}\label{sec:rw-state-surveillance}

\noindent State surveillance presents an immediate and pervasive threat to protest movements in Uganda, with the Ugandan State employing various surveillance strategies to monitor and control dissent~\cite{privacyinternational2015}. Beyond Uganda, prior scholarship has shown the impact of state surveillance on activists, with Bertrand~\cite{bertrand2018fear} and Monoghan and Walby~\cite{monaghan2017surveillance} researching environmental activists in Canada opposing critical infrastructure projects like new oil pipelines. In particular, Bertrand~\cite{bertrand2018fear} explored how targeted surveillance, backed by the implicit threat of state violence, shaped the behaviour of activists who challenged dominant economic, technological or cultural modes. Further, the authors of~\cite{sanches2020under} highlighted how activists under surveillance employ various coping strategies, including compartmentalising their lives to build layers of protection. They showed how belonging to a group or organisation may, on the one hand, expose individual activists to surveillance because of increased visibility while, on the other, provide a security culture with methods to elude surveillance. More broadly, studies on BLM protesters in the US highlighted particular surveillance concerns~\cite{Rosenbloom22}, while~\cite{CHI:BSCU21} and~\cite{CHIEA:WadBruFie21} studied the security advice given to participants in BLM protests, noting the focus on advice relating to circumventing surveillance. 

In a case study of Russian activists, Lokot~\cite{lokot2018safe} found that activists often combined public-facing social media accounts and encrypted communication channels to avoid surveillance. Albu~\cite{albu2023managing}
noted how activists' reliance on social media and digital tools for communication, mobilisation and organisation could open them up to state surveillance. Michaelsen~\cite{michaelsen2020resisting} highlighted how states might leverage their control over surveillance capabilities to monitor internet activity and track communication within activist networks. The authors of~\cite{sanches2020under} discussed how this could lead to the identification and targeting of activists, fostering a climate of fear and uncertainty, ultimately hindering their ability to organise and mobilise. A growing body of research has documented strategies employed by activists to circumvent state surveillance and censorship. In South Africa, activists responded to data-driven surveillance by developing ``pedagogies of resistance'', including education about state surveillance tactics and building solidarity networks~\cite{choudry2018activists}. Duncan~\cite{duncan2018activist}
compared activist responses to state surveillance in South Africa, the UK and Mauritius, showing how knowledge-sharing across borders helped develop effective counter-surveillance strategies. 

\subsection{Protests, Protesters and Related Security}\label{sec:rw-protesters-security-practices}

\noindent Some information security research has explored the experiences of protesters within contentious political contexts. Concretely, the authors of~\cite{USENIX:ABJM21} showed how protesters in Hong Kong aimed for secure, discreet communication and anonymity in their organisation of the Anti-ELAB protests. Here, anonymous polls in large group chats on Telegram were seen to achieve two security goals. While anonymity was seen to protect participants from potential risks associated with their protest activities, it also fostered a sense of shared ownership over chosen tactics, achieving `buy-in' from a large group of protesters. Further, the study exemplified how anonymity was aligned with a decentralised structure of protest organising in higher-risk environments where traditional leadership hierarchies can be vulnerable to targeted repression by authorities. In~\cite{SP:DSKB21} the authors demonstrated how protesters in the 2018-2019 Sudan revolution sought to protect themselves and each other in the event of compromise (usually arrest). In particular, features that allowed protesters to detect compromise were favoured (e.g.~live tracking applications), along with the ability to delete messages in group chats and to remove group members who were suspected to have been compromised. In~\cite{EUROCRYPT:ACDJ26}, the authors show how abductions were used for intelligence gathering during the anti-Finance Bill protests that erupted across Kenya in 2024. Here, the monitoring of protesters by Kenyan security forces led protesters to develop distinct security practices to try to protect themselves against targeted abductions; practices that were rooted in collective forms of security.  The author of~\cite{CHI:Guntrum24} also showed how the anti-military protests in Myanmar brought to the fore a series of security concerns for protesters in a politically contentious context, including the need for secure offline applications and specific application controls to protect information during physical compromises. In the context of climate activism, the authors of~\cite{USENIX:BrouJenAlb25} highlighted how decisions about security and technology use within the UK climate movement were often rooted in virtues that had little or nothing to do with security. 

Broader HCI scholarship highlights the centrality of digital technology in protest movements and their activities, while drawing out specific security implications, e.g.~\cite{CHI:dKMRW19,grinko2022nationalizing,CSCW:RAMRWW16}. The significance of security technology in organising, mobilising and amplifying protests is, however, highlighted in research beyond information security and HCI. For example, in~\cite{TrereReclaiming2015,Trere20} the author highlighted the extensive use of WhatsApp for ``backstage activism'' in the Mexican \emph{\#YoSoy132} movement in 2012-2013. Others have focused on the significance of WhatsApp in the 2013 Gezi protests in Turkey~\cite{ICS:ValVac18} and in the 2012 Occupy Nigeria protests~\cite{ICS:GHAC21}. 

Thus, technological tools are an integral part of many protest activities, as exemplified in the literature cited here, while protesters rely, in many ways, on the security of this technology for their protection. However, in both the 2019-2020 Anti-ELAB protests and the 2018-2019 Sudanese revolution, security research --~\cite{USENIX:ABJM21} and~\cite{DaffallaSecurity2021,SP:DSKB21}, respectively -- demonstrated the significance of both the social context and social relations within protest settings. For example, in~\cite{DaffallaSecurity2021} the author highlighted how the success and safety of a protest movement hinged on the ability of others to navigate security. In~\cite{USENIX:ABJM21} the authors showed how collectivity, social organising and shared security goals shaped protesters' security. In both contexts, the large-scale nature of the protests were significant to the security practices of protesters.

\paragraph{Research gaps.} Prior work highlights how larger-scale settings enable a degree of `security in numbers'. In contrast, the small scale of the Ugandan anti-EACOP context requires us to think quite differently about scale, where the individual protester is highly visible to an immediate adversary. Further, although some studies move beyond considering the security afforded protesters through their use of technology, e.g.~during deliberate internet shutdowns~\cite{SP:DSKB21,CHI:Guntrum24} or limited internet access~\cite{USENIX:ABJM21,CTA-RSA:ABJM21,USENIX:MccJenTal23}, we observe how the financial burden of internet connectivity remains largely unexamined. This is significant when considering the Ugandan context. With internet penetration at only 22\%~\cite{datareportalDigital2026} and data costs one of the highest in the world~\cite{surfshark2023Digital}, secure communication is a luxury that many cannot afford. Existing security-driven work on the needs of protesters and activists also highlights the collective nature of security practices within such settings, e.g.~\cite{USENIX:ABJM21,SP:DSKB21,CHI:Guntrum24}. Yet, the case of Ugandan anti-EACOP protesters points to autonomous decision-making (also seen in~\cite{USENIX:BrouJenAlb25}), where security is rooted in individual practices while situated within a collective programme of activism. Conducting our work \emph{in situ} enabled us to observe interactions \emph{as they happen} and engage participants in their environment. This provides insights for future information security research especially in a setting that has not previously been studied with respect to security.

\section{Research Design}\label{sec:research-design}

\noindent Focusing on a relatively small community of climate protesters in Uganda shaped our study design. Our primary concern was to ensure that participants could engage in the study in ways that did not further heighten their security risks, while we also considered fieldworker safety (\Cref{sec:ethics}).

\subsection{Fieldwork}\label{sec:methods}

\noindent One author (hereafter: \emph{the fieldworker}) conducted fieldwork in Kampala, Uganda's capital, over five weeks in July and August 2024. The fieldwork coincided with significant events, including the \emph{\#MarchtoParliament} protests on 23 July and two anti-EACOP protests on 9 and 26 August. During the same period, the offices of one of the organisations, with which some participants were affiliated, were raided and their devices taken. While this was done by unknown individuals, participants suspected State security operatives. Thus, the research was conducted in a context marked by heightened security among protesters, which led to rich insights into distinct security concerns. Yet, it also meant that it was not safe for the fieldworker to conduct participant observation at protest sites or to engage with protesters during and immediately after their arrest, although participant observation in those settings had originally been one of the planned research methods. 

Interviews were conducted in locations that ensured both participant and fieldworker safety, including NGO offices and public spaces, restaurants with private rooms (often used for private meetings in Kampala). In addition to formal interviews (\Cref{sec:interviews}) the fieldworker also engaged with participants informally. This included meet-ups and follow-up conversations over secure messaging (typically WhatsApp, which they preferred and already had installed on their devices), through which the fieldworker also established rapport with participants. It also meant that the fieldworker was sometimes informed about some activities being planned that only a small number of protesters knew in advance. Insights from these engagements were captured in note-form and brought into the analysis (\Cref{sec:data-analysis}). We refrain from detailing the specifics of such engagements for safety reasons.

\subsection{\emph{In-Situ} Interviews}\label{sec:interviews} 

Engaging participants in their \emph{naturally occurring settings}~\cite{brewer2000ethnography} through \emph{in-situ} interviews allows for greater depth, richness of meaning, contextual adequacy and comprehension in comparison to less immersive forms of interviewing, e.g.~remote interviews~\cite{gillham2005research,johnson2021beyond,lobe2022systematic}. These qualities are critical when studying sensitive topics and in contexts where surveillance is particularly prominent. We interviewed 13 participants (\Cref{tab:participants}), of whom 11 were frontline protesters\footnote{We use the term `frontline' to refer to protesters who have experience of confrontations with security forces during protests (defined in~\cite{USENIX:ABJM21}).} who had participated in at least one protest and been arrested or detained for their protest involvement. Anti-EACOP protesters are a small subset of climate activists engaged in protesting the EACOP, through forms of public demonstration to pressure the EACOP shareholders, including the Ugandan Government, TotalEnergies and CNOOC, to abandon the project. Two participants were climate activists who worked closely with the frontline protesters and provided legal and welfare support. We first recruited participants using purposive sampling~\cite{tongco2007purposive}, which involved reaching out to potential participants based on their involvement in the anti-EACOP protests. This approach was guided by the fieldworker's existing knowledge of the climate activist and advocacy landscape in Uganda. After the initial purposive sampling, we used snowball sampling by asking recruited participants for recommendations about other participants. The recruitment process was thus iterative and in line with our interpretivist position, avoiding a pre-determined `sample size'~\cite{braun2020saturation}. Rather, we focused on achieving depth in our interviews to allow fo the development of meaningful patterns throughout the dataset. 

\begin{table}[h!]
	{\centering 
		\caption{Overview of interview participants.}
		\begin{tabular}{cccc}
			\toprule
			(A) ID & (B) Occupation & (C) Role & (D) Duration\\
			\midrule
			P1 & Other & Other & 120 min\\ 
			P2 & Student & Core member & 60 min\\
			P3 & Student & Core member & 60 min\\
			P4 & Student  & Other &60 min\\
			P5 & Student & Core member & 60 min\\
			P6 & Unemployed & Other & 60 min\\
			P7 & Other & Other & 90 min\\
			P8 & Student & Core member & 120 min\\
			P9 & Other & Other & 45 min\\
			P10 & Unemployed & Other & 60 min\\
			P11 & Student & Other & 60 min\\
			P12 & Student & Other & 60 min\\
			P13 & Student & Core member & 90 min\\
			\bottomrule
		\end{tabular}\par
		\label{tab:participants}
	}
	\begin{tablenotes}
		\small
		\item\scriptsize\emph{Notes: (A) participant ID; (B) participant occupation at the time of the interview with some marked as 'other' for anonymity; (C) the position that the participant held in their activist group; (D) approximate duration of the interview in minutes}
	\end{tablenotes}{}
\end{table}

\noindent We used a topic guide to explore the threats facing anti-EACOP protesters, their fears and protection strategies. The guide was developed within the research team (see `researcher positionalities' in~\Cref{sec:data-analysis}), while it was piloted within the research setting. This led to some minor refinements and it informed how we engaged participants through the fieldwork. Further, the interview process was iterative; we moved between data collection and analysis, adjusting questions in successive interviews as new insights emerged. The interviews were conducted conversationally to facilitate an engaged discussion. 12 out of 13 interviews were conducted in English as most participants and the fieldworker were proficient in the language. One interview was conducted in Luganda, a widely-spoken local dialect familiar to both the participant and fieldworker. This interview was translated to English by the fieldworker before analysis. The choice of language was left to the participants. All interviews were audio-recorded with the explicit, informed consent of each participant (\Cref{sec:ethics}). The fieldworker also captured detailed notes. 

\subsection{Data Analysis}\label{sec:data-analysis}

\noindent We analysed our data inductively through several iterative coding cycles. Interview recordings were manually transcribed by the fieldworker and combined with interview notes. We were guided by Braun and Clarke's approach to reflexive thematic analysis~\cite{braun2006,braun2019reflecting}, with the fieldworker carrying out the first cycle of inductive coding, after familiarising herself with the full dataset. This was done manually and involved annotating the data, constructing higher-level categories and descriptive summaries of such categorisation. We ensured to continuously compare new data with previously collected data and constructed codes~\cite{GlaserStrauss99}, ensuring an iterative analysis. 

The second stage of the analysis involved both authors. The categories and associated descriptions and codes were extensively discussed in collaborative sessions. Here, the constructed categories were reflexively questioned, refined and revised, while interpretative summaries were created to ensure analytical agreement. Through this analysis process, the constructed categories were considered in relation to the evolving context of anti-EACOP protests, which is in line with prior work adopting this analytical approach (e.g.~\cite{USENIX:BrouJenAlb25,RobinsonJensen2025}).\footnote{Examples of our reflexive coding table can be found \href{https://doi.org/10.17637/rh.32305794}{here}.} 

\paragraph{Researcher Positionalities.} Our positionalities shaped how we conducted the research and interpreted the data. As noted by Hammersley and Atkinson in their seminal work~\cite{HammersleyAtkinson19}, social research is not conducted in a vacuum; it is part of the social world being studied, while also shaping it. The fieldworker is a Ugandan with experience in democracy promotion and oil sector governance. While not a climate activist, her familiarity with the country's political landscape provided a framework to understand the context in which protests were staged. The research team also involved a researcher who uses ethnography to study information security in higher-risk contexts. The research was thus designed -- and analysed -- from our individual and collective positions~\cite{braun2022starting}.

\section{Findings}\label{sec:findings}

\noindent We present our findings in line with the themes constructed through the reflexive thematic analysis (\Cref{sec:data-analysis}). In~\Cref{sec:structure-anti-eacop}, we set out how anti-EACOP protesters organised themselves. We focus on adversaries in~\Cref{sec:adversaries}, while compromise and participants' fears of compromise are outlined in~\Cref{sec:fears}. In~\Cref{sec:countering-surveillance} we show how anti-EACOP protesters aimed to protect information about themselves -- healthcare, living arrangements, movements -- and their activism to evade state surveillance.

\subsection{Structures of Anti-EACOP Protests}\label{sec:structure-anti-eacop}

\noindent The first major anti-EACOP protest was staged in October 2022, where a group of university students marched to the EU offices in Kampala, to deliver a petition supporting an EU Parliament resolution to halt the EACOP project (\Cref{sec:activism-uganda}). Nine of them were arrested and charged with `common nuisance'. Their devices were captured and some protesters noticed they had been tampered with upon return. While protests have continued, they have remained small in numbers, with an average of 20 protesters in most protests and often involving the same groups of protesters. Generally, anti-EACOP protests are designed to appear spontaneous, with protesters gathering from different locations to minimise the risk of security forces intervening and suppressing the action.

Anti-EACOP protests comprise a small number of interlinked and largely homogeneous groups. They are mainly staged by protesters affiliated with three predominantly student groupings that identify as movements. These movements initially emerged as student activists across different universities, mostly in Kampala, and coalesced around a common goal to stop EACOP developments. These loose movements operated in a fluid and informal manner and were heavily influenced by existing social networks at universities. Some participants held, or had previously held, leadership roles in their respective universities and had leveraged these positions to motivate others to join the cause. Others were drawn to anti-EACOP protests after participating in activism on other issues in Uganda; for example, protests over rising commodity prices and university fees. For many participants in our study, the bonds formed during their shared experiences of participating in other protests together had laid the ground for collectively staging anti-EACOP protests. 

Each movement had a core group of protesters who usually assumed greater responsibility for the planning of protests than others. These core members were individuals who had a strong commitment to the protests, demonstrated by their repeated participation, significant influence and trust within their movement, considerable efforts to recruit others into the movement and a clear passion for the cause. The different movements, together with other protesters, formed a broader network of protesters who were known to each other and who often collaborated in staging protests.  

The activities and `membership' of specific movements remained loosely defined, allowing flexible participation and decentralised decision-making in protest planning. Protesters participated in protests in their individual capacity, flowing across movements and freely joining protests planned by any movement. Thus, participants engaged in actions by their personal choice, not due to their affiliation with a specific movement. The movements relied on distributed networks of coordinators responsible for specific tasks necessary for staging each protest. These tasks, which included media and legal liaisons, operations and welfare management, were held interchangeably by different members for each protest and were tied to the specific protest more than a movement.

Typically, protests were organised in a similar manner. A core group of four to seven protesters -- often those with coordination roles -- led the protest planning. Planning involved in-person meetings to agree a date, time and location for an upcoming protest and distribute tasks among members. Resources were evaluated and strategies for their distribution agreed. Security aspects discussed included communication protocols for informing other protesters, escape routes in the event of police interception and safekeeping of participant devices during a protest. The group also discussed which and how media personnel would be invited at the last minute, to limit the risk of information leakage to authorities. Each protest was typically led by a particular movement, although different protesters from other movements occasionally participated in protest planning, strengthening collaboration across multiple movements. P13 explained how the collaboration among the movements was considered an advantage: \textit{``If they arrest guys from this movement, then guys from the other one will be available to plan something.''} At times when some protesters had been arrested or were keeping a low profile after having staged a protest, protesters from another movement would lead the planning of a successive protest to maintain momentum. In addition to cross-movement collaboration, some protesters also engaged with some NGOs in the climate and environmental conservation space on their advocacy work around EACOP development.

To onboard new members into the movements, protesters would aim to recruit others in person. Existing protesters assumed the responsibility of engaging others in supporting the anti-EACOP cause. P8 explained how core members in their movement would try to recruit at least \textit{``five''} members, ideally those they could meet in person and account for. For protests, protesters were typically responsible for passing on protest information to members they recruited. 

\subsection{Identifying Adversaries and Their Threats}\label{sec:adversaries}

\noindent Ugandan police forces were considered the immediate adversary due to participants' frequent experiences of arrests and protest disruption. All 11 protesters had been arrested at least once during a protest, with many having experienced multiple arrests (\Cref{sec:fears}). Arrests were conducted by uniformed police, which reinforced their status as the main adversary and enforcers of state interests, including suppressing dissent. The police who usually arrested them appeared to operate in coordination with more advanced state security forces, suggesting a level of operational influence: \textit{``The policemen work to please their bosses, the state security''} (P13). Participants spoke of a stratified structure, viewing these other state security agencies as having advanced technological resources to engage in digital and physical surveillance of protesters beyond those of the regular police. Participants also remarked on the particularly brutal policing approaches of the oil and gas protection directorate of the police, specifically created to safeguard the country's oil and gas resources and investments.

\subsubsection{Threats from Foreign Interests}\label{sec:foreign-threats}

Participants stressed that actions of the security agencies were driven by broader EACOP development interests. Some participants highlighted the potential collusion between the Ugandan Government and the oil companies -- the French multinational TotalEnergies and the Chinese state-owned CNOOC -- involved in the EACOP project. For example, P2, P8 and P12 reported being visited in prison by an individual who identified themself as a TotalEnergies employee. P12 explained: \textit{``Whenever there is an arrest, he comes to prison and the police officers bring him to talk to us.''} P8 shared that the same person had been seen with Ugandan police officers during the interception of one of their protests, suggesting that TotalEnergies was complicit in assisting the police in quelling protests. The same person was mentioned by all three participants in individual interviews. They had been urged to cease their anti-EACOP protests (and offered money or a job to do so), asked for their parents' contacts and offered money to cater for their welfare needs while in prison. P1, on the other hand, expressed a strong conviction that \textit{``the China Government is involved in [our] surveillance''}, attributing this to the protesters' opposition to the interests of the Chinese state-owned CNOOC. P1 had no concrete evidence but cited the Chinese Government's reputation for, in their view, extensive surveillance practices. Thus, to the participants, such foreign interests were not merely stakeholders in the EACOP project but key actors within their adversarial landscape.

\subsubsection{Adversarial Intimidation Tactics}\label{sec:intimidation}

\noindent Participants described several experiences where they, or their families, had received direct threats due to their involvement in anti-EACOP protests. Participants explained how they frequently received phone calls where an unknown caller relayed threats aimed at coercing them into abandoning protesting. While some protesters suspected that threatening calls came from state intelligence officers (as the callers spoke English with Ugandan accents), others thought they were oil company representatives. P1, P2, P4, P12 and P13 each recounted incidents where they were contacted by unknown individuals, where the caller ID and contact number were not visible. For example, P12, who had been in prison for nearly a month following a protest, shared how they started to receive threatening phone calls after their release. These calls warned them that the caller knew their whereabouts and urged them to stop protesting. Threats were also issued in person. P3 shared that once upon delivering a petition to Parliament, they were held by plain-clothed security officers and told: \textit{``We have seen you protesting before and one day we will take you to a place where you will not be seen again.''}

Ugandan security agencies extended these threats to protesters' families. P2 and P3 shared how their parents had received threatening calls from unknown individuals, urging them to get their child to cease their involvement in anti-EACOP activities. Participants stressed how such threats were aimed at leveraging social pressure to curb the protesters' engagement in anti-EACOP protests. P3 shared how their father had decided to stop paying their university fees because they deemed their participation in protests to be anti-government and, therefore, exposing themselves and their family to retaliation from the state. As a security measure, many participants protected both their own and their relatives' phone numbers, names and locations to reduce the risk of surveillance and intimidation by intelligence services.

Threats to Ugandan anti-EACOP protesters were thus often direct, as exemplified here in the form of intimidation, while other forms of indirect or passive threats were observed in different surveillance practices by the Ugandan intelligence agencies (\Cref{sec:fears}). Such overt and covert threats led participants to focus their security on personal and activism-related information (\Cref{sec:countering-surveillance}).

\subsection{Protesters' Fears of Compromise}\label{sec:fears}

\noindent Participants' involvement in protesting the EACOP development made them highly visible to the Ugandan security forces. All participants talked about their fears and experiences of surveillance, and how this impacted their lives beyond protesting (\Cref{sec:surveillance-fears}); their fears related to the prospect of being abducted, assaulted or even killed (\Cref{sec:arrests}). 

\subsubsection{Compromise during (Arbitrary) Arrests}\label{sec:arrests}

All the protesters had experiences of arrest (\Cref{sec:adversaries}). Even though arrests were common during anti-EACOP protests, participants feared the potential consequences of being arrested. This included the immediate risk of physical violence during an arrest and detention, but was often grounded in the potential (immediate and longer-term) consequences of information compromise for themselves, the protest activities and other protesters. Arrests were seen as a way for security forces to disrupt current and future protest activities. When arrested, protesters had to disclose personal information, including their full identification details, residence and occupation details. This raised concerns over the compromise of information related to group membership, protest planning and activities (past and future), funding sources and the whereabouts of other protesters. P10 voiced how arrests could lead to future tracking: \textit{``When you are arrested, they ask you for your number and when you list it, police will hand it to the security people responsible for tracking phones.''} 

Participants also voiced fears over the use of torture to coerce those detained to reveal critical information pertaining to protest plans. They feared that the compromise of such information could directly lead to abductions: \textit{``Once they get your information, they will raid you in a numberless car and they will take people to torture chambers''} (P8). Some participants shared how they would give false information as a form of protection, believing that providing false personal details would offer some level of protection from surveillance and tracking after their release. For example, as explained by P5: \textit{``In my statement, I do not put my exact phone number. I lie about it also. I lie about where I stay or I just give a general area.''} Other protesters feared trumped-up charges: 

\begin{quote}
    \textit{``They can trump up charges like terrorism. They force you to open the house and then they plant an explosive and then use this to trump a charge of terrorism and for that you will be produced in military court. It is something that has happened to other activists.''} (P8)
\end{quote}

\noindent This led many participants to hide their residential information, sometimes also from fellow protesters (\Cref{sec:countering-surveillance}). 

Participants also lived with the constant fear of being arbitrarily detained, tortured or killed by Ugandan security forces, with P10 relaying how they had been arbitrarily detained in relation to their wider political activism, i.e.~not anti-EACOP protests. Other participants had not experienced arbitrary arrest, yet their fears were founded in their experiences of threats and the experiences of others. Some gave the example of Stephen Kwikiriza, a Ugandan environmental activist in the EACOP-affected region, who had been abducted by plain-clothed officers, assumed from the Ugandan military, tortured and detained for six days in June 2024~\cite{Reuters:detained2024}. Kwikiriza had, like many study participants, received threats about his work documenting the environmental impacts of oil developments in the Albertine region. This precedent led many participants to fear that their activism could ultimately result in them being killed: \textit{``My fear is death because I know activists in Uganda can be abducted and killed''} (P2). This was underscored by P1: \textit{``You can even die.''}

\subsubsection{Perceptions of Surveillance}\label{sec:surveillance-fears}

All participants assumed that they were under extensive forms of surveillance and recounted different experiences that solidified this assumption. For example, some participants expressed concerns that their devices had been compromised and were being monitored: \textit{``I think the Government might be bugging the protesters' phones [\ldots] because they often get to know their [protesters'] plans and interrupt them''} (P9). Participants' suspicion of surveillance was grounded in specific experiences. For example, P9 noted: 

\begin{quote}
	\textit{``Sometimes you are talking, and you hear a strange noise in your phone, and when you ask those security experts, they tell you that the phone is tapped and that is how we know. Like me I have changed a phone three times because of that. Of course for the first few months the phone is fine and after that the same noise comes back, then you know that it has been tapped again so you change again.''}  
\end{quote}
	
\noindent P9 was able to source a new phone when they suspected that their device had been compromised, while others could not. Some protesters presumed a tight collaborative system between the Uganda Communications Commission (UCC), the national communications regulator and state security: \textit{``The security agents deal with UCC to understand your movements''} (P8). P8 explained that the UCC facilitated phone tapping and required service providers to share call records for anyone of interest to state security agencies. 

P2 was concerned that their phone conversations were being constantly monitored by state security agencies, who they believed had access to their personal information and whereabouts. They reflected that \textit{``[state security] use phone tapping and also some way, I am not sure, but they find a way to find your location''}. While most participants assumed that they were being tracked through their devices, others also highlighted how their movements were being closely observed by state security. P5 shared how their neighbours had reported that unknown individuals carrying a photograph of P5 had visited their communal residence, asking if P5 resided there. Others, e.g.~P1, recounted how they, after being released from prison, had noticed a consistent pattern where the same vehicle appeared in a fixed spot near their home in the evenings. Beyond such experiences of physical monitoring, participants highlighted how \textit{``security agencies like ISO [Internal Security Organisation] create files on all of us''} (P9) to facilitate continuous monitoring. P9 expanded that when \textit{``they [police] arrested us last time [\ldots] they had files with all our names and they were telling us about different anti-EACOP activities and meetings we had had and the people we had met with.''}

\subsection{Autonomous Security in Collective Action}\label{sec:countering-surveillance}

\noindent Protesters collectively planned and staged protest activities in close-knit affinity groups, yet their security practices were grounded in autonomous decision-making. While some security practices were discussed among protesters and agreed on their viability, individuals retained the autonomy to tailor actions based on their own preferences and needs. We draw out how participants' security practices shaped their daily routines and practices to protect information related to themselves and their activism (\Cref{sec:everyday-activism}), while showing how protesters compartmentalised within groups to protect against protest-specific information leakage (\Cref{sec:compartmentalising}).

\subsubsection{Location and Identity Information}\label{sec:everyday-activism}

Participants' security practices extended beyond the planning and staging of anti-EACOP protests. Their fears about the potential consequences of compromise meant that they adopted a state of perpetual vigilance, which led them to integrate extensive information-protective measures into their lives. In particular, location and residential information dominated participants' security practices in our analysis. To protect against perceived surveillance tactics by Ugandan security forces (\Cref{sec:surveillance-fears}), participants adopted strategies such as frequently changing their living arrangements, choosing densely populated areas and maintaining a geographical distance from family members and fellow protesters and other relational ties. P5, whose previous neighbour had reported unknown persons asking questions about and showing a photo of P5, stressed how this had been the cue for them to \textit{``shift''}: \textit{``immediately you know that you have to shift.''} Some protesters refrained from disclosing their residential address; this meant keeping their living arrangements from friends, family and members of their movement: \textit{``I relocated and nobody knows where I stay. Now I do not know where anyone stays''} (P12). 

Participants shared how they hid location-specific details when using public services such as healthcare facilities, where some (e.g.~P2) would not list their correct address or phone number. Anti-EACOP protesters also typically resided in low-cost housing such as university hostels, even while they were not attending the specific university. P3 noted how they considered staying in a hostel as \textit{``safe''} because \textit{``we are so many, even if they wanted to pick me up, it would be hard to know which room I am in.''} Participants also avoided signing rental agreements, instead, asking friends to register tenancies on their behalf. They also refrained from paying water and power bills with their own mobile money numbers so as to protect what they considered sensitive information about themselves. Other location-specific security practices involved not turning on location settings on devices to, in their view, limit the risk of being tracked by security forces. P3 reflected that \textit{``I also switch my location app off in the settings, I never use it.''} 

Participants who kept two phones (\Cref{sec:tools}) usually had both a smartphone and a non-smartphone, and would use only the latter during specific, critical periods as they believed that it could not be used to track their locations. P7 explained how \textit{``they keep another phone like a Nokia, a `ka torchi',\footnote{\emph{Ka torchi} is a Ugandan colloquialism for a small non-smartphone.} to reduce the location tracking.''} Participants also refrained from staying at their usual residences, particularly the night before a protest. Protecting location information was especially critical before and after protest activities when participants were `laying low'. Participants also spoke about concealing their movements by, for example, not relying on one or the same mode of transport for their daily commutes. They exemplified how they would take two or three different boda bodas\footnote{Boda bodas are motorcycle taxis and common in East Africa.} to reach their destination: \textit{``I also do not see my family very often and when I go, I take many routes. I take two taxis and like three bodas until I reach home''} (P5). P13 explained how they would \textit{``not take the first boda that agrees to take me, we bargain, and I pretend I don't have enough money and then take the next one.''} Participants reasoned that changing routes and transport provided them with a degree of protection as it meant their movements could not be predicted and information about their whereabouts harder to obtain.

For many participants, concealment of their identity was a feature of their day-to-day lives. When P6's neighbours mentioned having seen P6 on TV being arrested in a protest, they lied and said it was not them: \textit{``I am staying in my mother's house so I don't want them spying on us.''} Relatively new to protesting, they were reluctant to admit involvement in the protests, fearing that their neighbours could be Government informants or sympathisers: \textit{``There are always spies so I have to be careful.''} While others, e.g.~P2 and P5, would adopt similar concealment approaches, they reasoned that their purpose was to protect against negative social repercussions for their families and to limit the risks of Ugandan security forces targeting them. 

\subsubsection{Compartmentalising Information Access}\label{sec:compartmentalising}

Keeping protest information from leaking was considered critical as leaked information could lead to pre-emptive arrests, disruption of planned protests and exposure of participants to violence. Ahead of planned protests, protesters would adopt extensive information security practices to reduce the risk of information compromise. For example, details of upcoming protest locations and times were typically sent the night before, through a tiered system, where each protester would communicate to those in their contact list, usually via WhatsApp if they could not inform them in person. Participants also shared how meeting times and locations would sometimes be altered at the last minute to protect such information which was considered critical to the staging of the protests. As noted by P2: \textit{``Most times the police intervene when we are close to the destination.''} 

While anti-EACOP protesters did not rely on secure messaging applications in their daily communication due to data costs, they recognised the benefits of encrypted communication. Thus, the sharing of information relating to upcoming protests was done through WhatsApp as they considered this to be the `most secure' way to distribute sensitive information. However, given that most protesters adopted a practice of regularly switching off data (due to data costs), such information would sometimes be received too late while some protesters were relying on out-of-date information. This tiered system also meant that some protesters in the last tier would receive the communication from different individuals. Despite these overlaps in information dissemination, participants explained that the tiered approach minimised the risk of compromise by limiting the number of protesters with early access to protest information, which was considered critical. 

An anti-EACOP protest planned in August 2024 underscored the importance of the compartmentalisation of information access across protester networks. \textit{Taxis} (mini vans) taking protesters to the protest site were intercepted by Ugandan police, arresting 47. The interception happened only minutes before arrival at the protest site and the agreed start time of the protest. This protest represented the largest ever mobilisation for an anti-EACOP protest, with over 50 protesters invited to participate, including many new protesters mobilised by local civil society organisations (CSOs). Some participants suggested that the information might have leaked due to the relatively higher number of protesters with whom the information was shared -- and given that many were new to protesting. Highlighting this tension, P13 noted: \textit{``I distance myself from civil society organisations [\dots] they are too much into proposals and policy. I do not share a lot of information to civil society. The Government has the capacity to infiltrate them.''}

\subsection{Identified Security Tools}\label{sec:tools} 

\noindent We explore the `security tools' adopted by anti-EACOP protests, with a focus on encryption (\Cref{sec:encryption}) and phones (\Cref{sec:phones}), highlighting how such practices were often grounded in their fears about the consequences of information and device compromise. Security practices among anti-EACOP protesters varied and can be mapped to participants' prior experiences of compromise. P1, who had been arrested several times in the last two years, described how their phone was always in \textit{``lock mode''}: 

\begin{quote}
	\textit{``Every time I change a phone, I set the lock mode before placing my SIM card in it so that I can know when my phone is being tracked [\ldots] I have got a notification before that someone has entered the serial number of my phone in a tracking software.''} 
\end{quote}

\noindent This experience, along with numerous arrests, led P1 to reason that they were under state surveillance. As a result, they adopted distinct -- and often extensive -- measures to protect their phones and information. For example, like P9 who had changed their phone thrice due to concerns about it being tapped, (\Cref{sec:surveillance-fears}), P1 changed their phone ``often'', calling this tactic a \textit{``good security practice''} because they believed it made it hard for the security forces to track them. Other participants explained that they had multiple SIM cards or used two phones, with some (e.g.~P5, P10, P12, P11 and P7) explaining that having two phones was an attempt to compartmentalise their lives; dedicating one phone primarily to their protest-related activities, although that distinction was described to not always be easy or clear. Further, P10 explained a nightly routine for securing their phones including switching them off and hiding them in different parts of their house. P5 had established a similar practice: \textit{``When I am going to sleep, I hide my two phones in case they break into my house and I am taken.''} This strategy was informed by the fear that they could be taken at any point -- also during the night. Others, however, considered that resetting the phone to its factory settings during an abduction, or at times when they were particularly concerned about being taken, would protect them as this \textit{``wipes out everything''} (P8). Thus, participants were more concerned about state security forces having access to their phone data than the potential consequences of them discovering that all data had been wiped.

Limiting access to phones was a driver for many participants' security practices. Most protesters relied on passwords, PINs or patterns to secure their phones. While P3 and P5 mentioned biometrics, P5 voiced concerns about the vulnerability of biometric authentication during an abduction: \textit{``On my phone, I do not use the fingerprint security feature or the face ID because I know that when I am kidnapped, they can forcefully open my phone.''} Protesters who received threatening calls, viewed changing SIM cards as a tactic to protect against such threats. P12 explained: \textit{``I keep changing phone numbers [\ldots] after two months I buy a new number. When I am called, I change a number.''} Although changing numbers did not stop the threatening calls, it made P12 feel more secure. Since phone numbers are linked to identities in Uganda, SIM cards would often be registered using other people's national IDs. P5 shared how they \textit{``got an old man in my friend's village to register for me a SIM card''} to avoid it being linked to P5. 

\subsubsection{Concealing for Protection}\label{sec:encryption}

Most participants in our study preferred using encrypted messaging applications over traditional SMS and phone calls, with WhatsApp being used significantly more frequently than Signal. However, most participants also spoke about not being able to afford continuous data required to use secure applications. For example, P3 noted: \textit{``Buying data all the time is expensive [\ldots] as soon as you buy it gets done.''} The high cost of data also restricted VPN use: \textit{``I use VPN because someone advised me that I can avoid being tagged and located but it takes up all my data so sometimes I switch it off''} (P2). Similarly, P8 explained: \textit{``If I send people messages and they remain on one tick for a while, it means they have no data, so I just call them.''} The prohibitive cost of data disrupted communication among anti-EACOP protesters, hindered coordination of protest actions and forced them to rely on less secure modes of communication. 

Some participants disguised their secure messaging applications by changing their icons to blend in with other applications. For example, P1 explained that they would hide certain applications in their gallery application icon to dissuade security officials from accessing sensitive communications such as conversations with fellow protesters, in the event of device capture. Others adopted practices such as saving contacts under pseudonyms and memorising phone numbers instead of storing them in their phones. P4 explained how they would not store their mother's and brother's numbers in their phone but \textit{``in my head and even the conversations are deleted [\ldots] you cannot get my mum's or my brother's numbers.''}

Protesters deleted messages from their phones to remove communication traces that security forces could use to get information about those they communicated with and protest activities. P5 deleted their conversations with family members immediately after each conversation. P2, P3, P4 and P5 considered switching off their phones for extended periods of time a protection mechanism. P4, for example, believed that by leaving their phone switched off, they could protect their parents when visiting them in \textit{``the village''}. P3 would usually activate a call-forwarding function to a non-existent number as they wanted \textit{``to control who can call me so I do not pick up calls from numbers that I do not know''}. 

\subsubsection{Protesting With or Without Phones}\label{sec:phones} 

Participants agreed that leaving phones at home would be the `correct' approach, yet this was not followed by everyone and remained an individual rather than a collective decision. Some protesters carried phones to protests, sometimes because they coordinated key aspects of the protest, other times out of personal preference. Yet, given the high risk of compromise during anti-EACOP protests, i.e.~arrests (\Cref{sec:arrests}), most participants left their devices at home. P8 shared how during their first arrest they had brought their phone that \textit{``had all the information on all our plans.''} Their phone had been taken at the police station and although they got it back when they were released \textit{``my phone number had been blocked.''} Losing their phone number disrupted their communications, isolated them from their networks and restricted their ability to access essential mobile money services.  

While leaving phones at home was seen as a way to limit device compromise during protests, participants highlighted how leaving their phones at home limited their actions. P8 noted that if the decision was made to change course \textit{``we cannot communicate to everyone, sometimes people miss out.''} In one protest, where many protesters left their phones behind, their ability to communicate had been significantly constrained. The police intercepted the leading vehicle with protesters. Since they were unable to communicate this to others, all vehicles were intercepted and most protesters detained. 

\section{Discussion}\label{sec:discussion}

\noindent The Ugandan State is heavily invested in the securitisation of the oil sector, indicated in part by the establishment of a dedicated police directorate (\Cref{sec:adversaries}) for safeguarding oil and gas resources and assets (\emph{``the oil police''}), the labelling of anti-EACOP protesters as `economic saboteurs' (\Cref{sec:activism-uganda}) and the suppression of civil society engagement in oil governance issues~\cite{hrw2023working}. In his seventh re-election campaign in 2025, Museveni highlighted the oil sector and the military as key achievements of his rule~\cite{oketch2025museveni}, signalling the importance of the sector for regime legitimacy. As such, any challenge to its development, including anti-EACOP protests, is framed as a security threat, prompting the surveillance of protesters and the use of extensive force and intimidation tactics. In addition to the interests of the Ugandan Government, international commercial interests are at stake, not least exemplified by how arrested anti-EACOP protesters were visited in prison by a TotalEnergies employee. Hence, while the anti-EACOP protests are small in scale, their potential impact, as perceived by the State, is significant -- rooted in commercial interests, regime legacy concerns and geopolitical relations, with France's TotalEnergies and Chinese state-owned company CNOOC as critical partners. 

Given the clandestine nature of protest planning and execution, coupled with the severe threat of arrest, violence, abductions and trumped-up charges, protesters operate under heightened information security conditions. In response to these threats, anti-EACOP protesters deployed adaptive strategies that reflect their specific security concerns. We discuss the implications for information security, before we conclude by setting out directions for future work in~\Cref{sec:conclusion}.

\subsection{Scale Matters}\label{sec:scale-matters}

\noindent Existing information security scholarship shows how the security practices by anti-government protesters hinge on more than security technologies (\Cref{sec:rw-protesters-security-practices}). They are grounded in the social and political contexts, and adversarial landscape, within which the protests are planned and staged, e.g.~\cite{USENIX:ABJM21,SP:DSKB21}. Our work shows how Ugandan anti-EACOP protesters adapted their security practices to \emph{their} specific context which focused on limiting access to information about future protest activities as a way to reduce the risk of police disruption and infiltration before the start of a protest. This was made possible because of their small-scale context, where availability to information could be limited in time and place, which created an environment where security practices rested on individual rather than collective decisions (\Cref{sec:tools}). 

Existing literature on the use and limitations of communications technology during protests largely focuses on mass protests such as BLM~\cite{CHI:BSCU21,Rosenbloom22}, the Arab Spring~\cite{US:AlsGuv15,OUP:HowHus13,moss2016transnational}, the Gezi protests~\cite{ICS:ValVac18}, the 15M movement~\cite{ICS:MicCas14} and the Occupy movement~\cite{ICS:Gerbaudo17}, to mention a few. Similarly,~\cite{USENIX:ABJM21},~\cite{EUROCRYPT:ACDJ26},~\cite{USENIX:BrouJenAlb25},~\cite{SP:DSKB21} and~\cite{CHI:Guntrum24} concerned larger-scale settings which fundamentally bound the ways in which activities were organised through a large number of more or less interlocking groups. In these settings, findings suggest a reliance on security technology (e.g.~secure messaging applications) for in-group communication, social media for wider (international) mobilisation (e.g.~diaspora as seen in~\cite{SP:DSKB21} and~\cite{CHI:Guntrum24}), large Telegram channels for collective decision-making~\cite{USENIX:ABJM21}, and conflicts over application selection~\cite{USENIX:BrouJenAlb25}. The authors of~\cite{USENIX:ABJM21} further note how protesters' choice of messaging applications largely depended on group size. Thus, while not explicitly stated, prior work highlights the significance of (also) considering scale when designing for the information security needs of protesters.

Our study points to the importance of protest scale in a different way. Their small scale meant that, by necessity, anti-EACOP protests in Uganda were characterised by secrecy and spontaneity (\Cref{sec:structure-anti-eacop}) as protesters could not rely on `security in numbers' or collective security approaches as observed in~\cite{USENIX:ABJM21,EUROCRYPT:ACDJ26,SP:DSKB21} for large-scale settings. Similarly, spreading the risk of protest organising across several individuals by either framing the protests as `leaderless'~\cite{USENIX:ABJM21} or adopting a `decentralised' structure across connected leaders~\cite{NMS:AzeHarZhe19} was not an achievable security goal for Ugandan anti-EACOP protesters. Instead, their small scale allowed for clandestine protest planning. This offered particular protection in the form of controlling access to future protest information to a select, trusted few, which was seen to limit the risks of infiltration, disruption, pre-emptive arrests and information compromise (\Cref{sec:compartmentalising}). This contrasts with larger-scale protests where organising relies on a broader network of protesters and, as a result, accepting that not everyone's identity can be verified~\cite{USENIX:ABJM21}. Our findings show -- exemplified by the 47 pre-emptive arrests in August 2024 -- how the opening up of the protest to wider participation left it exposed to information leakage and, as a result, pre-emptive arrests. As others have observed~\cite{USENIX:BrouJenAlb25}, this underscores a key security tension within close-knit activist networks: the strive for secrecy to protect information and group members, on the one hand, and the need for openness to mobilise and grow the movement, on the other.

The small scale of the protests was fundamental to the decisions participants made about their security and, thus, underpins the findings we report in~\Cref{sec:findings}. The participants in our study were highly visible to the adversary -- \emph{as individuals} -- regardless of their status in the movement. This led to lifestyle-altering security practices to protect against state surveillance and capture, including moving into temporary accommodation, altering travel movements and cutting communication with and deleting contacts of fellow protesters and family members \Cref{sec:everyday-activism}. This contrasts with observations made in larger-scale protests, where those considered to be leading organisers are targeted for their protest involvement. This was seen in the anti-Finance Bill protests in Kenya in 2024, for example, as reported in~\cite{EUROCRYPT:ACDJ26}. On the flip side, the small scale allowed for a hierarchical and clandestine structure where sensitive information about upcoming protests could be limited to a select few and only communicated close to the time of the action to protect against preemptive arrests \Cref{sec:compartmentalising}. This gives rise to information security practices grounded in small and highly trusted activist networks. 

The protests being small-scale shaped the research design. The small number of protesters, combined with the heightened security at the time of the fieldwork, meant that the fieldworker refrained from engaging participants during active protests (\Cref{sec:methods}). It also meant that most participants knew each other and had participated in the same protests. Thus, while small-scale protests require distinct organising, we observe how research designs need to consider scale -- to both protect participants and researchers given the increased risk of identification and, by extension, targeting.

\subsection{Autonomous Security in Groups}\label{sec:autonomous-security}

\noindent Protests are founded in collective action, while existing work points to the significance of `connective action'~\cite{ICS:BenSeg12} and `connective leadership'~\cite{NMS:AzeHarZhe19} for large-scale protest organising and mobilising. Ideas of `collectivity' and `connectivity' through extended information and organising networks have also been observed in security scholarship on larger-scale protests, e.g.~\cite{USENIX:ABJM21, USENIX:BrouJenAlb25,SP:DSKB21}. For example, the authors of~\cite{SP:DSKB21} showed how networks of local groups and leadership facilitated country-wide information-sharing during internet shutdown. We found that, albeit at a small scale, Ugandan anti-EACOP protesters relied on similar structures for cascading protest-relevant information, through core members, and the recruitment of new members (\Cref{sec:structure-anti-eacop}). Yet, collective forms of information security were notably absent within the context of Ugandan anti-EACOP protests, where the protection of information about protest activity and group membership and counter-surveillance strategies relied upon protesters' autonomous decision-making. In~\cite{USENIX:ABJM21}, the authors showed how Anti-ELAB protesters' individual security notions were negotiated in line with those of their group, while protesters adopted the protective strategies collectively decided for their group. This was not the case for the participants in our study.

We found that for Ugandan anti-EACOP protesters, `security in numbers' was not a security goal. Their security practices were grounded in their own decisions rather than negotiated between and collectively decided upon by group members. For example, the choice to use WhatsApp for some protest-related communication was based on most protesters having (intermittent) access to WhatsApp, not a collective decision to rely on this application. Those who could not be reached via WhatsApp would receive relevant information through other channels, e.g.~phone calls, SMS or in person. The communication of upcoming protest activities was done through what might be called hierarchical tree structures, where protesters received and shared information with others on their protester contact list (\Cref{sec:compartmentalising}). Yet, this approach had developed organically within groups grounded in practical considerations such as protesters' individual contact lists for fellow protesters. This also meant that some protesters would receive the same information from different protesters. Thus, we observe that, for the participants in our study, time-bound \emph{availability} of information was a key security notion (\Cref{sec:temporalities}). To enable this, they relied on autonomous, rather than shared, security practices. 

\subsection{Countering (State) Surveillance}\label{sec:countering-state-surveillance}

\noindent Prior work, e.g.~\cite{michaelsen2020resisting}, highlights the expansive reach of state surveillance through control of communications infrastructures, while studies like~\cite{lokot2018safe} and~\cite{albu2023managing} demonstrate the duality of visibility and secrecy in technology tools and platforms used by activists. Others have shown how activists develop counter-surveillance strategies through knowledge-sharing and solidarity networks, e.g.~\cite{choudry2018activists,duncan2018activist}. We found shared counter-surveillance efforts to be notably absent among anti-EACOP protesters as they relied on individual security. 

Our findings present a different dimension of state surveillance and the security practices adopted by protesters. Ugandan anti-EACOP protesters' hyper-vigilance towards surveillance was conditioned on the tangible consequences of living in a repressive political environment rather than from the technological capabilities of the Ugandan security forces -- or their own `tools'. While the participants in our study assumed the Ugandan State's surveillance capabilities to be significant, their counter-surveillance practices were driven by anticipating abductions and arrests, i.e.~compromise. Since the actions of anti-EACOP protesters are seen to threaten high-stakes economic and commercial interests and international alliances (\Cref{sec:adversaries}) the risks attached to their surveillance come with equally high stakes, with participants reasoning that abductions, torture and death might be potential consequences of their activism. Thus, while we are not the first to highlight the impact of state surveillance on protest movements, the small scale, limited resources and societal context that anti-EACOP protesters had to navigate, led to distinct security practices. Such practices not only contribute to a broader conception of state surveillance of resistance movements, but show how protesters make decisions about protecting their information.

\subsection{Temporalities of Security}\label{sec:temporalities}

\noindent Our work shows the centrality of time-bound information availability for anti-EACOP protest activities, with protesters developing information-access and communication strategies that changed with time. In the hours leading up to protests, core members released information about upcoming activities to other protesters; this often meant that many protesters were only alerted to upcoming activities less than 12 hours before they were to take place (\Cref{sec:compartmentalising}). Thus, we observe that `time to protest' was a significant security consideration for anti-EACOP protesters. Participants also shared how they would change their protective strategies in the time after being released from prison following an arrest (\Cref{sec:tools}), as they feared state surveillance (\Cref{sec:surveillance-fears}). This led them to switch off their phones, spend time in a different location and move residence with the aim of re-establishing some form of anonymity. Further, during arrest, participants noted how they would provide false identity information which they assumed offered some degree of protection in the form of pseudonymity (\Cref{sec:arrests}). Temporality was thus an integral aspect of security for participants, where considerations related to protection \emph{before} and \emph{during} protests and protection \emph{during} and \emph{after} arrests fundamentally shaped their security. This suggests the need to model the temporal dynamics of protection strategies at different stages of protest planning, staging and recovering. This is also shown in~\cite{USENIX:ABJM21}, where the authors consider changing access requirements across time and place for Anti-ELAB protesters within small affinity groups. 

\subsection{Resource Constraints}\label{sec:resource-constraints}

\noindent Much prior work has focused on the role of communications technology in facilitating protest organising. While media literature points to the significance of this technology in large-scale protests, e.g.~\cite{Castells15,TrereReclaiming2015}, security scholarship points to distinct practices reliant on security technology as we discuss in~\Cref{sec:rw-protesters-security-practices}. Yet, the Ugandan context showed a different dimension of the reliance on security technology: data costs. While participants expressed a preference for secure messaging such as WhatsApp over SMS or phone calls, their inability to afford uninterrupted data meant that they were unable to use these tools consistently -- or rely on them for their protection (\Cref{sec:tools}). Thus, contrary to the assumption made by Signal when they stopped supporting encrypted SMS/MMS in 2015 -- \textit{``[t]here are certainly places where data is not accessible, but those are also mostly places where smartphones are equally inaccessible''}~\cite{Signal2015} -- we show that although access to data was limited in the setting we studied, participants had smartphones, making encrypted SMS particularly attractive.

While this dimension remains underdeveloped in the security literature, others have pointed to the limits of relying on internet-enabled communications technology in resource-constrained contexts and during political unrest. For example, the authors of~\cite{USENIX:MccJenTal23} showed how unreliable internet access and rising data costs in post-conflict Lebanon resulted in limited access to secure communications, while the authors of~\cite{SP:DSKB21} pointed to the reliance on analogue strategies such as the use of SMS and regular phone calls during the five-week internet shutdown in the Sudanese revolution. Our work underscores the gap in understanding security practices in low-resource contexts. Indeed, the reliance on secure messaging assumes a degree of affordable and reliable access to an internet connection. Yet, this does not reflect the realities of many living in regions like Uganda, and broadly Sub-Saharan Africa. Consequently, protesters, in often politically contentious settings, are typically forced to resort to less secure communication channels, such as SMS or phone calls, often leaving them increasingly exposed to state surveillance and targeting.

Some attempts have been made to cater to settings with no or limited internet. Yet, these attempts have mainly focused on large-scale settings, e.g.~the development of mesh network applications that offer communication capabilities over Bluetooth -- a technology often being promoted for use during times of political unrest or internet shutdowns (see e.g.~Bridgefy~\cite{Bridgefy2022,Bridgefy2021}). As others have shown, the use of this technology in adversarial contexts has not been successful while it also presents distinct security risks. In~\cite{SP:DSKB21}, the authors noted how the adoption of mesh networking applications in the Sudanese revolution was \emph{``unsuccessful''}. More critically, in~\cite{CTA-RSA:ABJM21} and~\cite{USENIX:AlbEikPat22}, the authors presented serious security vulnerabilities in the Bridgefy application, which promoted itself as \emph{``the protest app''}. The small-scale nature of the Ugandan anti-EACOP protests presents further challenges for the design of offline solutions, with existing mesh network options being both insecure and relying on mass adoption. 

\section{Conclusion: Future Work}\label{sec:conclusion}

\noindent The Ugandan anti-EACOP context sheds light on a series of open questions for security researchers. 

Frequent arrests and fears of abduction -- and pervasive surveillance -- led to heightened concerns about information compromise among protesters. Future work might consider specific forms of access control that allow for information protection \emph{during} a compromise, i.e.~as it happens.\footnote{Anti-ELAB protesters voiced a similar concern of wanting protection during a compromise~\cite{USENIX:ABJM21}.} Participants referred to the practice of doing a `factory reset' that would `wipe everything', suggesting that information protection outweighed concerns over loss of information.

Information about upcoming protest activity was shared through hierarchical structures to limit information leakage. Future work might consider modelling temporality in the design of security technology, recognising that security needs are dynamic and change over time and place, such as before, during and after an event (e.g.~protest, arrest).

Resource constraints meant that protesters could not rely on secure messaging for the sharing of time-critical communications as access to data could not be assumed. Future work might consider the potential for encrypted voice and SMS services in settings where data cannot be assumed but where people have smartphones, as was the case for participants.

The small scale of anti-EACOP protests make protesters highly visible to a powerful adversary that has the means, and is willing to use these means, to compromise them through arrests, interrogations and forms of surveillance. Future work might consider how to design for such a high-risk context. 

Researching small-scale protests requires an approach that is sensitive to these conditions. We were not able to do planned observations due to the heightened security during the time of the research while we `only' spent five weeks in the field setting. Future work might consider longer-term fieldwork with smaller-scale protest and activist networks.

\paragraph{Limitations.} Our fieldwork coincided with two anti-EACOP protests in August 2024, the aftermath of a prior anti-EACOP protest held in June 2024, and a major anti-corruption protest in July 2024. This politically charged and volatile context heightened the risks to both the fieldworker and participants, thereby limiting on-the-ground observations of common aspects to protests in Uganda such as arrest, police interactions and protest planning (\Cref{sec:methods}). Thus, while participant observation, initially considered a central method for us, would have enriched the insights gained from the fieldwork, it was not possible. We consider this a future research direction. Additionally, the secretive nature of protest activities and the sensitivity of security-related questions occasionally led to participant apprehension, somewhat constraining data collection. Engaging protesters over a longer period of time, including through ethnographic fieldwork, has the potential to overcome such a limitation. Although interviewing 13 participants is significant given that the wider group of active anti-EACOP protesters comprises fewer than 30 individuals, the small size of this network means each protester's security experience is uniquely shaped by their circumstances. We tried to overcome this limitation through our reflexive analysis by drawing out observed patterns of threats, fears and protective mechanisms. We also note, as with all ethnographic work, that our findings are particular to the context under study. While the initial coding cycle was done by the fieldworker, subsequent cycles and iterative analysis was carried out between the authors. This aligns with our ethical standpoint as only the fieldworker can have access to the dataset. 

\section{Acknowledgments} 

This work would not exist without the many contributions from anti-EACOP protesters and supporters in Uganda, who so generously gave of their time to speak to us and trusted us with their experiences. The research of Mbabazi was supported by the UKRI as part of the Centre for Doctoral Training in Cyber Security for the Everyday at Royal Holloway, University of London (EP/S021817/1). The research of Jensen was supported by the UKRI as part of the Social Foundations of Cryptography project (EP/X017524/1).

\section{Ethical Considerations}\label{sec:ethics}

\noindent Ethical approval for all research components was obtained from the authors' institutional Research Ethics Committee (REC) prior to the start of the research. The research was classified as \emph{high risk} and a full risk assessment was carried out and approved by the Health \& Safety (H\&S) office of the authors' institution. However, given the evolving protest context during the fieldwork, additional and dynamic safety protocols were established to ensure the safety of both the fieldworker and the participants when engaging in the research. 

Only by being in place could we gain insights into the security practices that these groups established for themselves -- and only by gaining such insights can we propose how we as security researchers might better serve groups that are otherwise hard-to-access and, thus, under-represented. We proceeded with this research and submission because the intended benefits outweighed the mitigated risks which we judged to be controllable and acceptable.

\paragraph{Participants.} We recognise that the participants in our study are in a sensitive context where revealing certain details could, in theory, be used against them by their adversaries. To minimise this risk, we adopted rigorous and reflexive ethical practices and safety protocols. Upon initial contact, participants received an information sheet which outlined the key objectives and scope of the study, as well as what participation would involve. 

Participation was voluntary and participants had full agency over the nature, frequency and extent of their participation. We took great care in ensuring that all participants were aware that they could withdraw from the study at any time without giving a reason (no participant withdrew their contributions). This information was contained in participant information sheets, while the fieldworker ensured to reiterate that participants should only share as much as they felt comfortable sharing before any engagement. In some instances, participants shared information that they specifically asked not to be included in any write-up from the study but decided to share for the purpose of enriching other aspects of the discussion. We, without exception, adhered to such wishes. 

An important feature of our fieldwork is also that some participants initiated further engagements to share additional information that they deemed useful to the research and, in so doing, exercised agency over their participation in the study. Participants' participation in some respects was a demonstration of their desire to document the conditions under which they practised security and make them visible to others.

Participants did not receive financial compensation, as we did not want participants to feel compelled to take part because of a financial incentive, and thus possibly accept increased risks (this is in line with the policy of our institution(s)). Audio recordings were made only with prior permission from each participant, and all interview participants consented to this. Some participants shared screenshots of messages and physically showed the fieldworker information on their mobile devices. We treated this data in the same way as any other research data obtained during the fieldwork, where we omitted/concealed any sensitive details, digitised and then destroyed any physical material, and securely stored this data. Data was anonymised with identifiable details removed, and importantly, participants provided informed consent not just at the start of any research engagement but continuously throughout the fieldwork. Consent was obtained verbally for data minimisation and cultural appropriateness purposes, and as established through our institutional ethics approval. We do not include participant demographic summaries to minimise the risk of de-anonymisation, given the potential to link quotes with individuals. Further, in a bid to protect participant details, our presentation and discussion of findings omit operational details that would directly endanger participants, and instead, we focus on broader patterns rather than potentially sensitive specifics. We have also omitted specific locations in the reporting of our findings given the context of heightened state surveillance. We strove to protect participants' anonymity throughout the research and treated their information confidentially at all stages. 

We brought our findings back to participants before submitting our work for review. The reason for this was two-fold: we wanted to share our findings as a form of reciprocity and, critically, to ensure that any information contained in any published work would not unknowingly or accidentally reveal aspects of their activities, organising or daily life they considered sensitive, detrimental to their work and potentially risky. Similar conversations took place during the fieldwork. No concerns about our work were raised.

\paragraph{Non-Participant Stakeholders.} We considered the potential risks to a wider group of stakeholders in the setting under study, while still keeping in mind the fundamental risks to participants from such stakeholders and recognising the power dynamics within this context. We deliberately only refer to non-participant stakeholders at an institutional level to mitigate harms to individuals. Specifically, the oil companies we mention in our work could in theory experience reputational damage; however, their involvement in the EACOP development is already widely publicised (also by the companies themselves). Further, we protect individual employees by only referring to the companies rather than identifying individuals, e.g. ``TotalEnergies employee''. Similarly, we do not identify individual members of the Ugandan security forces, but refer to the units at an institutional level.

\paragraph{Publication.} Participants trusted us with their experiences as a way to contribute to a better understanding of their protests and the repressive conditions under which they organise. We view this work as an opportunity to amplify their voices within the information security community. Not publishing this work would therefore be a disservice to them.It is also important to highlight that participants engaged with the work in an iterative manner, often prompting contact with the fieldworker to share additional information and invite her to engagements, reflecting the agency they exercised and deliberate nature of their involvement and interest in the research. Similarly, in some instances, participants shared information that they specifically asked not to be included in any write-up from the study but decided to share for the purpose of enriching other aspects of the discussion. We, without exception, adhered to such wishes. 

\paragraph{Fieldworker Wellbeing.} To minimise risks to the fieldworker, a safety protocol was established within the research team. This was developed in consultation with the fieldworker's institutional H\&S office and involved regular check-ins and ongoing discussions about necessary adaptations in response to the evolving context, and we took advice from participants. Check-ins served both as a vehicle to discuss fieldwork progress and for the fieldworker to reflect on experiences in the field. We also had a course of action in the case any check-ins were missed, but we never had to enact this.

The research did not present psychological stress to the fieldworker, who has a distinct understanding of the specific setting and many years of experience engaging with civil society and resistance groups in Uganda (while not a member of such groups). The research team holds extensive ethnographic experience, also from higher-risk settings, while the fieldworker's institution provides wellbeing services to support researchers. Yet, our internal debriefings did not lead to the involvement of the wellbeing team. Lastly, as we also note in \Cref{sec:methods}, the fieldworker did not participate in or directly observe protests to reduce exposure to potential harm. 


\section{Open Science}\label{sec:open-science}

Given the sensitive nature of this work and our commitment to protect the participants we work with, we do not to share the full dataset, including interview transcripts and notes as well as field notes. This data includes specific experiences and perspectives that cannot be abstracted to a point where individuals may not be identified from the data, even if specific identifiers were to be removed from the dataset. This decision is in line with established protection practices in qualitative social research, especially in a context where making such data available would put participants at greater risk. 

Building rapport with research participants is critical in ethnographic research as it relies on mutual trust, often developed over long(er) periods of engagement and rooted in confidentiality with the participants. In order for us to to adhere to our ethical principles and responsible research practices, we respected this confidentiality at all stages -- including at the dissemination stage. Further, for the safety of participants, we only provide example excerpts from interviews in our work. We use pseudonymised and carefully curated quotes to support our findings and to ensure that the participants' voices remain present in our work. We also compartmentalised within the research team so that only the fieldworker had (and continues to have) access to the full dataset. 

\paragraph{Data Management.} It is also worth highlighting that before commencing the research, we established a detailed data management plan which considers both accidental and forced data leakage, and is based on the principle of data minimisation. This plan was submitted and approved with our full ethics application and we continue to adhere to it. All data was digitised and securely stored before the fieldworker left the field setting, to avoid accidental and/or forced data leakage while crossing international borders.

All interview transcripts and field notes were anonymised with identifiable details removed during the transcription process and stored in an encrypted folder on the fieldworker's full disk-encrypted laptop, which was kept in a secure location throughout the fieldwork. The data collected, specifically anonymised observation notes and anonymised interview transcripts, will be retained in the fieldworker's academic institution's research data archive for 10 years for safeguarding in line with institutional policies and our ethics approval, after which they will be securely destroyed. Note that this institution is situated in a country with strong legal protections and Uganda is not in a strong position vis-a-vis this country. 

\medskip

\noindent \emph{We make our topic guide, participant information sheet and examples of our reflexive thematic analysis table available \href{https://doi.org/10.17637/rh.32305794}{here}.}


\bibliographystyle{plain}
\bibliography{local}

\end{document}